\documentclass[conference]{IEEEtran}
\IEEEoverridecommandlockouts
\usepackage{cite}
\usepackage{amsmath,amssymb,amsfonts}
\usepackage{algorithmic}
\usepackage{graphicx}
\usepackage{textcomp}
\usepackage{xcolor}
\usepackage{booktabs}
\usepackage{array}
\usepackage{tabularx}
\usepackage{makecell}
\usepackage{amssymb}
\usepackage{graphicx}
\usepackage{subcaption}
\usepackage{float}
\def\BibTeX{{\rm B\kern-.05em{\sc i\kern-.025em b}\kern-.08em
    T\kern-.1667em\lower.7ex\hbox{E}\kern-.125emX}}
\begin{document}

\title{Sensor-Level Fault Diagnosis for Automotive Software Validation Using Large Language Models}

\author{\IEEEauthorblockN{1\textsuperscript{st}  Mohammad Abboush*}
\IEEEauthorblockA{\textit{Technische Universität Clausthal} \\
Clausthal-Zellerfeld, Germany \\
mohammad.abboush@tu-clausthal.de}
*Corresponding author
~\\
\and
\IEEEauthorblockN{2\textsuperscript{nd} Hamza Ouarrad}
\IEEEauthorblockA{\textit{Technische Universität Clausthal} \\
Clausthal-Zellerfeld, Germany \\
hamza.ouarrad@tu-clausthal.de }

~\\
\and
\IEEEauthorblockN{3\textsuperscript{rd} Andreas Rausch}
\IEEEauthorblockA{\textit{Technische Universität Clausthal} \\
Clausthal-Zellerfeld, Germany \\
andreas.rausch@tu-clausthal.de}
~\\
}

\maketitle

\begin{abstract}
The pre-series validation of automotive software on hardware-in-the-loop (HIL) platforms produces large volumes of multivariate sensor recordings whose assessment against functional safety requirements exceeds what manual review can sustain at campaign scale. Threshold-based tooling reports that a deviation has occurred but neither identifies its nature nor locates its source, while data-driven classifiers, although accurate, rely on large labelled datasets and return opaque decisions that sit uneasily with the traceability demanded by ISO 26262. This study examines whether open-source instruction-tuned large language models (LLMs), given a textual description of sensor behaviour, can serve as data-efficient and interpretable engines for fault detection and diagnosis inside the validation loop. A two-phase framework is proposed: automated requirement checking on a dSPACE real-time platform first isolates the recordings that violate a safety requirement, and only these are inspected, with sliding windows of the signals reduced to statistical, relational, and contextual descriptors, embedded in a fixed prompt, and mapped by a 4-bit low-rank-adapted LLM to a fault location accompanied by a written justification. Four model families ranging from two to eight billion parameters were adapted and tested on a gasoline-engine case study spanning six injected fault classes. The smallest model matched the largest at 81.6\% accuracy, whereas a comparably sized model failed to converge, indicating that diagnostic competence under task-specific adaptation follows convergence rather than parameter count, with the entire adapt-and-evaluate cycle fitting on a single commodity accelerator.
\end{abstract}


\section{Introduction}
The development of automotive software systems is governed by stringent functional safety requirements that are validated across the V-model life cycle in accordance with the ISO 26262 standard \cite{url}. Depending on the maturity of the system under test, these activities are distributed across the X-in-the-Loop testing stages \cite{abboush2024virtual} and culminate, before series production, in virtual and real test drives that validate the integrated vehicle against its safety requirements \cite{abboush2022hardware}. Hardware-in-the-loop (HIL) simulation, which couples physical electronic control units (ECUs) and selected components with high-fidelity real-time plant models, has become a central enabler of this process, since it allows reproducible, safe and cost-efficient execution of demanding scenarios \cite{mihalic2022hil}. A single run of an advanced driver-assistance or powertrain function may record hundreds of signals at one to ten kilohertz over several hours, producing gigabytes of multivariate, strongly nonlinear and multi-mode time-series data, and a full campaign can comprise thousands of such runs. Manual review at this scale is cognitively intractable, while current tool-based analysis with manually defined thresholds \cite{theissler2013anomaly} only flags that an anomaly has occurred, cannot detect unanticipated failure modes, and characterises neither its type nor its severity.

Once abnormal behaviour is flagged, locating the fault remains difficult, because anomalies often propagate through several interconnected components before becoming observable in the recorded signals \cite{marashi2021identification}.


Data-driven fault detection and diagnosis (FDD) has therefore gained traction \cite{dai2013knowledge}, with machine learning (ML) and deep learning (DL) achieving strong automotive fault-classification performance at low engineering cost \cite{abboush2022hybrid}. Convolutional and recurrent architectures, in particular, reach high accuracy on time-series benchmarks \cite{lei2020review}, yet they exhibit two limitations that are severe in this domain. They require large, balanced, and well-labelled datasets that conflict with the scarcity and class imbalance of real validation data, and their opaque representations offer little of the human-interpretable explanation needed to justify a diagnosis \cite{abboush2025xai}. This lack of transparency complicates ISO 26262 certification, weakens engineers’ trust, and impedes debugging and transfer across vehicle variants, a difficulty that is aggravated by simultaneous faults \cite{valdez2020cascading}.

In parallel, large language models (LLMs) can perform reasoning and classification tasks directly from natural-language instructions, often zero-shot or few-shot, while keeping their parameters fixed, with chain-of-thought prompting and in-context learning improving performance across domains \cite{sahoo2024prompt}. Recent work has begun serialising sensor measurements into text for diagnosis \cite{gruver2023timeseries}, raising the question of whether open-source LLMs, paired with a signal-to-text representation, can serve as data-efficient and explainable FDD engines. However, the empirical comparison of open-source LLM families for sensor-level FDD under realistic automotive conditions remains limited: existing LLM-based diagnosis largely targets software logs and incident reports rather than multivariate physical signals or functional-safety constraints \cite{zhang2025llm}, and the behaviour of parameter-efficient fine-tuning, such as LoRA, has not been characterised for this task across model families. In particular, open-source LLMs have not been examined for fault detection and classification within the real-time HIL validation of automotive software systems across internal-combustion engines under high-fidelity simulation.

To address these gaps, this article proposes and evaluates an LLM-based framework for sensor-level FDD embedded directly in the ISO 26262 validation workflow. The framework operates in two consecutive phases. First, the functional safety requirements are validated on the HIL platform using an industrial dSPACE tool chain, which executes the test scenarios and automatically identifies the recordings in which a requirement has been violated. Second, only these violated recordings are forwarded for intelligent inspection: sliding windows of the multivariate sensor series are converted into structured natural-language descriptors capturing statistical patterns, abrupt changes, and context such as vehicle speed and operating mode, and passed to open-source instruction-tuned LLMs prompted to jointly output a fault label and class. This coupling concentrates the diagnostic effort on safety-relevant deviations and aligns the analysis with the requirement-driven structure of ISO 26262. We study four model families of differing scale and design, namely Gemma 2 2B-it, Qwen2.5 3B, Llama 3.1 8B and Mistral 7B, under LoRA fine-tuning, and validate the framework on two HIL case studies, a gasoline-engine system, benchmarked against strong deep-learning baselines.
The main contributions of this work are as follows:
\begin{itemize}
\item A unified framework bridging multivariate automotive sensor signals and LLM reasoning through a signal-to-text representation and structured prompting for fault detection and classification during real-time validation.
\item A two-phase pipeline integrated with the ISO 26262 process: safety requirements are first validated on an industrial dSPACE HIL platform, and only the recordings of violated requirements are forwarded to the LLM, coupling standards-compliant validation with requirement-targeted diagnosis.
\item To the best of our knowledge, the first systematic benchmarking of multiple open-source instruction-tuned LLMs for sensor-level FDD in the HIL validation context under a shared LoRA adaptation regime and a common evaluation protocol.
\item Validation on an industrial gasoline-engine case study executed on a high-fidelity real-time HIL platform, with transfer to further propulsion architectures.
\end{itemize}

The remainder of this article is organised as follows. Section 2 reviews related work; Section 3 details the proposed framework; Section 4 describes the experimental set-up and case studies; Section 5 presents and discusses the results; and Section 6 concludes and outlines future work.

\section{Related work}

Automotive FDD has progressed from shallow classifiers towards deep architectures that learn discriminative representations directly from raw multivariate signals. Convolutional and densely connected pipelines have been applied to sensor-fault detection and prognostic health-index forecasting in autonomous-driving stacks, with Safavi et al. \cite{safavi2021multi} reaching 99.84\% detection accuracy on the Audi A2D2 dataset but restricting evaluation to four canonical fault classes injected through a static distribution model. Recurrent architectures have dominated diagnostics in electrified powertrains, exemplified by the LSTM model of Kaplan et al. \cite{kaplan2021fault}, which reduced the mean absolute percentage error from 10.13\% to 2.06\% over a shallow ANN baseline, although fault injection was performed only at the Simulink level. Hybrid and ensemble formulations have broadened coverage and robustness: a two-class/one-class ensemble attaining a 77\% F2-score on OBD-II recordings under known and unseen faults \cite{theissler2017detecting}, a multi-label LSTM and Random Forest ensemble augmented by a GRU denoising autoencoder reaching 99.43\% accuracy on concurrent sensor faults under HIL-generated noise \cite{abboush2023ensemble}, and a Random Forest classifier coupled with fault-tolerant control on a TruckMaker-based HIL platform \cite{raveendran2020brake}; complementary reviews note that many algorithms simplify actuator dynamics and neglect parameter drift \cite{pietrowski2024fault}. Across these works, accuracy is high, but evaluation remains confined to narrow fault catalogues, predominantly single-fault scenarios and single-platform validation, while the dependence on large labelled datasets and the opacity of learned representations limit both generalisation and the interpretability required for ISO 26262 certification.

A complementary line targets functional safety verification directly. At
circuit level, dual-view graph neural networks with gradient-based
attribution reach up to 99.5\% accuracy while halving fault-injection
data and meeting ASIL-D metrics at reduced cell area
\cite{sun2025iso}, and graph convolutional criticality
regression generalises across circuits but degrades on heterogeneous
designs \cite{das2024graph}. At system level, a CI-enabled HIL framework
combining a denoising autoencoder, LSTM classification, and k-means
clustering reaches a 91.85\% F1-score \cite{abboush2025advancing},
predictive fault management for ADAS controllers improves diagnostic
coverage by 25\% under ASIL-D constraints \cite{abdul2026ai}, and
integrated fault-injection-and-classification frameworks attain 88--92\%
accuracy \cite{nasri2026integrated}. These approaches, however, operate
on circuit graphs rather than physical sensor signals or exclude the
learning component from the safety path, and their explanations remain
feature-attribution maps rather than human-readable engineering
justifications.

The reasoning and instruction-following capabilities of large language models (LLMs) have prompted growing interest in their use for automotive diagnosis. A textual strand classifies large fleets of fault-symptom claims \cite{pavlopoulos2024automotive} and generates evidence-grounded reports through CTGAN, RAG, and GPT-4o pipelines \cite{mahale2025automated}, achieving high readability but only moderate factual accuracy. A retrieval strand embeds structured knowledge into diagnostic assistants, including RAG coupled with knowledge graphs for new energy vehicles \cite{zhang2025large1}, distributed on-vehicle diagnosis under network constraints \cite{chen2026personalized}, and HIL-GPT, whose domain-adapted compact models outperform larger counterparts in HIL test-sequence retrieval \cite{feng2025smarter}. 

In aggregate, automotive LLM research has concentrated on textual artefacts rather than multivariate physical signals under functional-safety constraints, rarely compares open-source model families, and has not quantified the benefit of LoRA over prompt-only adaptation. No prior work systematically benchmarks open-source instruction-tuned LLMs for sensor-level fault detection and classification within real-time HIL validation across internal combustion. The present framework addresses this gap: windowed sensor signals are converted into structured natural-language descriptors, LLM-based inspection is embedded in the two-phase ISO 26262 workflow so that only requirement-violating recordings are diagnosed, and several open-source LLMs are evaluated under LoRA regimes against strong deep-learning baselines, jointly producing a fault label and classification of Location\_ID.

\section{Methodology}

\begin{figure*}[!t]
    \centering
    \includegraphics[width=0.90\textwidth]{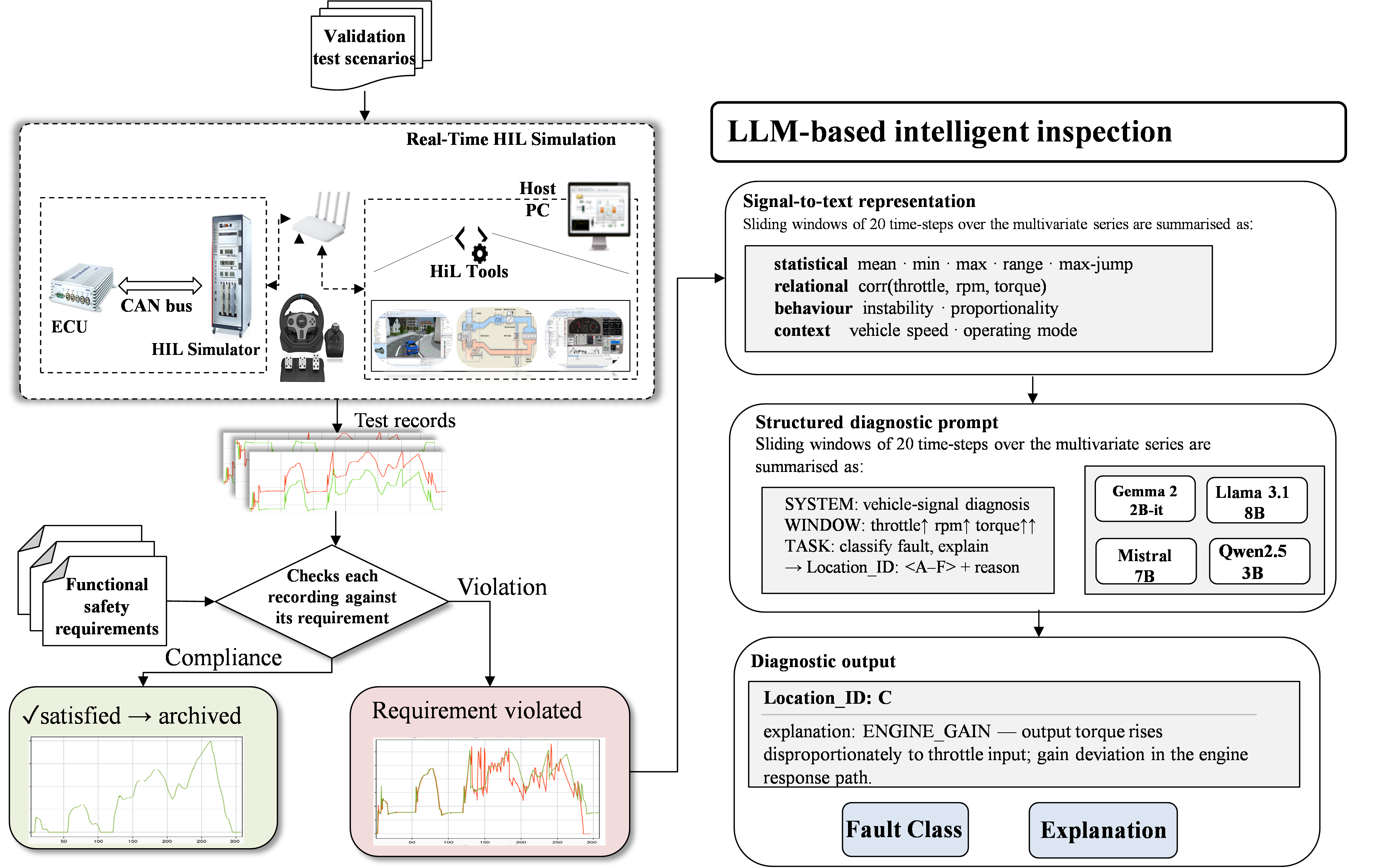}
    \caption{Overview of the proposed LLM-based intelligent inspection framework.}
    \label{fig:framework}
\end{figure*}

The proposed framework for the sensor-level detection and diagnosis of faults during the real-time validation of automotive software systems is illustrated in Figure~\ref{fig:framework} as two consecutive phases, i.e., real-time HIL validation with requirement checking, and LLM-based intelligent inspection. The first phase executes the functional safety requirements on the HIL platform and isolates the recordings in which a requirement has been violated. The second phase forwards only these violated recordings to a language model, which receives a textual summary of the corresponding sensor window and returns the fault type and its location. Separating the two phases reflects the requirement-driven structure of ISO~26262: the analysis effort is concentrated on safety-relevant deviations rather than on the full body of recorded data, and the diagnostic output is tied directly to the requirement that triggered it.

\subsection{Phase 1: Real-time HIL validation and requirement checking}
The first phase relies on an established industrial tool chain rather than
on any new method. The plant and environment models are executed on a
dSPACE real-time simulator; the control logic runs on a physical or
emulated ECU connected through the CAN bus, and the simulator, the ECU,
and the host PC exchange sensor, actuator, and control variables in real
time, as shown in the left part of Figure~\ref{fig:framework}. This
set-up, described in detail in our earlier work~\cite{abboush2024virtual}, produces multivariate recordings that capture the
system behaviour before, during, and after the activation of an injected
fault.

Each recording is evaluated against the functional safety requirements of
the corresponding function, in line with the system-level integration,
testing, and safety-validation activities of ISO~26262 Part~4 (clauses
4-8 and 4-9). An example of such a requirement for the engine control
function is: "In the presence of implausible accelerator pedal or
throttle position signals, the SUT shall limit the effective engine
torque to the driver demand derived from the last plausible pedal
position, and the engine speed shall not exceed the demanded value by
more than 10 per cent for longer than 500~ms." Compliant recordings are
archived without further processing, whereas recordings that breach their
requirement are flagged as violations and forwarded to the inspection
phase. Within the proposed framework, this phase therefore serves a
single purpose: it acts as a filter that keeps the inspection cost of
Phase~2 proportional to the number of safety-relevant deviations rather
than to the total test duration.

\subsection{Phase 2: LLM-based intelligent inspection}

The inspection phase answers the three questions that a requirement check alone cannot resolve, namely the nature, the location, and the probable root cause of the detected fault. A pre-trained LLM is employed for this purpose, since such models perform classification and reasoning directly from natural-language instructions and produce a human-readable justification alongside the prediction. The phase comprises a signal-to-text representation, a structured diagnostic prompt, the LLM with its adaptation regime, and the diagnostic output.

\subsubsection{Signal-to-text representation} 

A language model cannot operate on raw numerical sequences. Long series of floating-point values exceed a practical prompt length and are poorly handled by token-level processing. Each violated recording is therefore segmented into sliding windows of fixed length, and every window is summarised by a compact set of descriptors that preserve the diagnostic content of the segment. Following the central block of Figure~\ref{fig:framework}, the descriptors are organised into four groups: statistical, relational, behavioural, and contextual.

For a window of $L$ consecutive samples of a signal $x$, the statistical descriptors comprise the mean, the minimum, the maximum, the range, and the largest one-step change,
\begin{equation}
\footnotesize
\begin{aligned}
\mu_x &= \frac{1}{L}\sum_{t=1}^{L} x_t, \quad m_x = \min_{t} x_t, \quad M_x = \max_{t} x_t, \\
r_x &= M_x - m_x, \quad j_x = \max_{1 \le t < L} \left| x_{t+1} - x_t \right|,
\end{aligned}
\label{eq:stats}
\end{equation}
where $x_{t}$ is the value of the signal at time step $t$ within the window, $\mu_{x}$ is the window mean, $m_{x}$ and $M_{x}$ are the window minimum and maximum, $r_{x}$ is the peak-to-peak range, and $j_{x}$ is the maximum jump between consecutive samples. The range and the maximum jump expose, respectively, the amplitude of the excursion and the abrupt transitions that often accompany an injected fault. The relational descriptors capture the coupling between physically related channels through the Pearson correlation

\begin{equation}
\footnotesize
\rho_{ab} = \frac{\sum_{t=1}^{L}\left(a_{t} - \mu_{a}\right)\left(b_{t} - \mu_{b}\right)}{\sqrt{\sum_{t=1}^{L}\left(a_{t} - \mu_{a}\right)^{2}}\;\sqrt{\sum_{t=1}^{L}\left(b_{t} - \mu_{b}\right)^{2}}},
\label{eq:corr}
\end{equation}
where $a$ and $b$ denote two signals, such as throttle position, engine speed, and torque, and $\rho_{ab} \in [-1, 1]$ quantifies the strength and sign of their linear dependence. A loss of the expected correlation between proportionally linked channels is a strong indicator of a fault in their common response path. The behavioural descriptors translate these relations into qualitative attributes, such as the instability of a signal and the proportionality between coupled signals, while the contextual descriptors record the operating situation, in particular the vehicle speed and the operating mode, so that a deviation is interpreted relative to the regime in which it occurred.

 \subsubsection{Structured diagnostic prompt}
 
The descriptors of a window are assembled into a structured prompt that conditions the model on the task, the evidence, and the required answer format. The prompt is organised into three parts. An instruction block defines the role of the model as an automotive signal-diagnosis assistant and states the task, i.e., to classify the fault. An input block lists the window descriptors together with the contextual flags, providing the evidence on which the diagnosis rests. An output block fixes the response schema, requiring the model to return a fault location identifier from the predefined set of candidate locations together with a short reasoning. Constraining the output to this schema enables automatic parsing of the predicted label and keeps the accompanying explanation grounded in the supplied evidence. The instruction template is fixed across all windows, whereas the input block is generated separately for each window from its descriptors.

\subsubsection{Open-source LLMs and adaptation regimes}

Four open-source instruction-tuned model families of differing scale and design are studied as the inspection engine, namely Gemma~2 2B-it, Qwen2.5 3B, Llama~3.1 8B, and Mistral 7B. Each model is a decoder-only Transformer that maps the tokenised prompt to a sequence of contextual representations through stacked self-attention and feed-forward blocks, and generates the answer autoregressively. 

The models are evaluated under LoRA fine-tuning regimes. 


The pre-trained weights are quantised to a 4-bit representation and kept frozen, and a pair of trainable low-rank matrices is attached to the linear projections of the attention and feed-forward layers. The adapted weight of such a layer is:

\begin{equation}
W' = W_{q} + \Delta W = W_{q} + \frac{\alpha}{r}\,B A,
\label{eq:qlora}
\end{equation}
where $W_{q}$ is the dequantised 4-bit base weight, $B \in \mathbb{R}^{d \times r}$ and $A \in \mathbb{R}^{r \times k}$ are the low-rank adapter matrices, $r \ll \min(d, k)$ is the adapter rank, and $\alpha$ is a scaling factor. Only $A$ and $B$ are optimised, while $W_{q}$ remains fixed, which reduces the number of trainable parameters and the memory footprint by orders of magnitude relative to full fine-tuning. The adapters are trained by minimising the token-level cross-entropy of the reference responses,
\begin{equation}
\mathcal{L}(A, B) = -\sum_{m=1}^{M} \log p_{\Theta'}\!\left(y_{m} \mid y_{<m}, P\right),
\label{eq:ce}
\end{equation}
where $y_{m}$ is the $m$-th token of the target response of length $M$, $y_{<m}$ are the preceding tokens, and $\Theta'$ denotes the model parameters with the active adapters. Because the base weights are shared and frozen, a separate lightweight adapter can be maintained for each model without duplicating the underlying network, which makes the comparison across the four families and the industrial case studies practical on a single accelerator.

\subsubsection{Diagnostic output}
The model returns a structured response containing the predicted fault location and the fault class, as shown in the lower right part of Figure~\ref{fig:framework}. The label supports the quantitative evaluation of the classification, while the explanation is examined for correctness and engineering usefulness. Because each prediction is accompanied by the reasoning on which it rests, the validation engineer can accept or reject the diagnosis on technical grounds rather than trusting an opaque label, which shortens the manual root cause analysis and supports the integration of the component into the certification workflow. The framework is applied to an industrial gasoline-engine case study modelled in high-fidelity real-time simulation; its design contains no engine-specific elements, so transfer to other propulsion architectures is a natural extension examined in future work.

\section{Case Study and Data Description}


\subsection{ASM gasoline engine case study and HIL platform}  
The framework is evaluated on the dSPACE ASM gasoline-engine system in a closed-loop HIL environment. The engine is treated not as an isolated plant but as part of an integrated vehicle system coupling the powertrain, vehicle dynamics, and driver and environment models, so that an injected abnormality can propagate through connected subsystems before becoming observable in the recorded signals. The engine model covers the air path, fuel system, piston engine, exhaust system, and cooling system, together with lower-level elements such as the intercooler, throttle valve, intake manifold, EGR path, turbocharger, and torque-generation components, making the case study well suited to studying how a local signal manipulation affects both the target subsystem and the surrounding vehicle behaviour.

\begin{figure}[!t]
    \centering
    \includegraphics[width=0.50\textwidth]{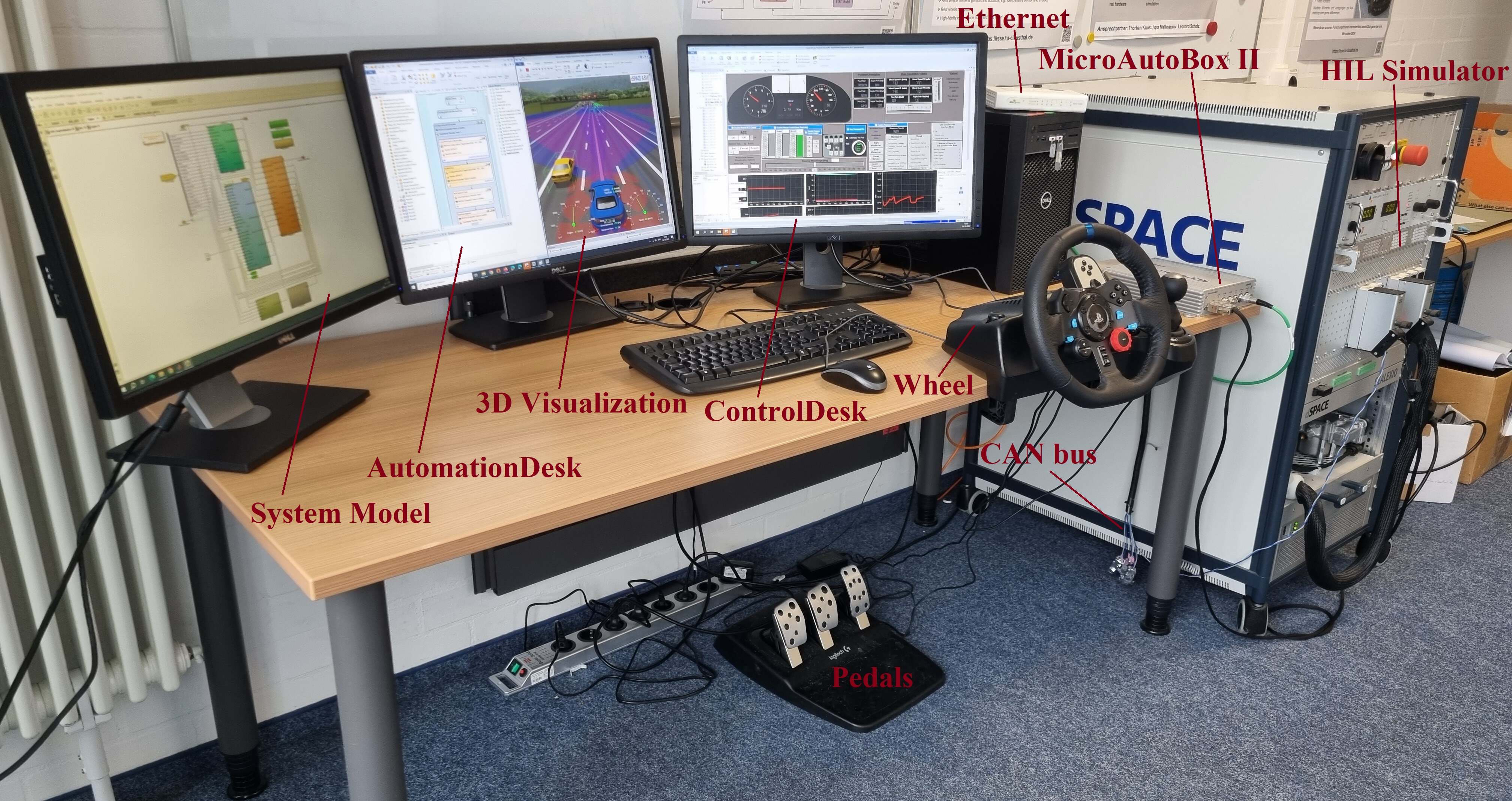}
    \caption{Experimental HIL simulation platform used for dataset generation and validation.}
    \label{fig:hil_setup1}
\end{figure}

The plant and controller models are developed in MATLAB/Simulink. The plant and environment models run on a dSPACE SCALEXIO real-time simulator, while the control logic is deployed to a dSPACE MicroAutoBox II emulating the target ECU in real time, with sensor, actuator, and control variables exchanged through standard interfaces under realistic timing constraints, as shown in \ref{fig:hil_setup1}. The controller supports online operation against the emulated ECU and offline operation against a SoftECU for pre-validation. Faults are introduced by controlled injection, so that nominal and abnormal behaviour are recorded under identical conditions, and the resulting recordings supply the violated cases diagnosed in the inspection phase.

\subsection{Signal acquisition and window-based representation} 
The recordings are stored as multivariate CSV files of synchronised measurements sampled at 0.5-second intervals. Six channels are retained because they characterise the engine and control-system behaviour relevant to the injected faults, namely engine speed, throttle position, effective engine torque, engine-out temperature, rail pressure, and fuel flow. Single-row inference on these signals proved unstable in preliminary testing, since an isolated sample conveys neither the short-term dynamics nor the inter-channel relationships that distinguish one fault from another. Each file is therefore segmented into overlapping windows of twenty time steps with a stride of three, so that every window spans approximately ten seconds of behaviour while successive windows preserve temporal continuity.


Each window is reduced to the compact descriptor set defined in Section 3. The statistical descriptors comprise the mean, the minimum, the maximum, the range, and the largest one-step jump of the selected signals, and capture both the amplitude of an excursion and the abrupt transitions that often accompany an injected fault. The relational descriptors are the correlations between throttle and torque, throttle and engine speed, and engine speed and torque, and expose the loss of the proportional coupling expected between these physically linked channels. Together with the Boolean contextual flags described below, these descriptors constitute the evidence from which the natural-language prompts are built, and they are computed identically for every window and every model.

\begin{table*}[htbp]
\centering
\caption{Training settings of the fine-tuned models.}
\label{tab:train_settings}
\begin{tabular}{lcccccccc}
\toprule
Model & Epochs & LR & Batch & Grad.\ accum. & Eff.\ batch & LoRA $r$ & $\alpha$ & Max len. \\
\midrule
Gemma 2 2B-it & 5 & $8\times10^{-5}$ & 1 & 8  & 8  & 8 & 16 & 768  \\
Qwen2.5 3B    & 6 & $1\times10^{-4}$ & 1 & 8  & 8  & 8 & 16 & 1024 \\
Llama 3.1 8B  & 2 & $3\times10^{-5}$ & 1 & 32 & 32 & 1 & 2  & 128  \\
Mistral 7B    & 5 & $8\times10^{-5}$ & 1 & 32 & 32 & 2 & 4  & 640  \\
\bottomrule
\end{tabular}
\end{table*}

\subsection{Prompt construction}

\begin{figure}[!t]
    \centering
    \includegraphics[width=0.50\textwidth]{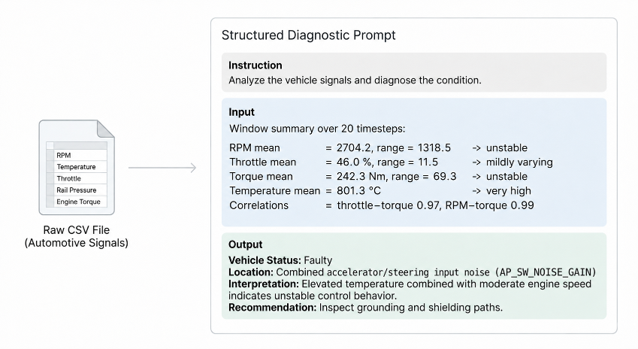}
    \caption{Structured diagnostic prompt used as input to the LLM.}
    \label{fig:hil_setup}
\end{figure}

The structured prompt exposes the sensor evidence to the language models in a deterministic form that is identical across the dataset. Rather than raw time series, each window is summarised into a numerical snapshot of its dominant operating behaviour and embedded into a fixed instruction-input-output schema. The instruction field states a single diagnostic task, i.e., analysing the vehicle signals and identifying a potential fault. The input field lists the snapshot values of engine speed, temperature, throttle position, rail pressure, and engine torque in physical units, together with the source-file identifier for traceability and Boolean state flags, such as over-temperature and wide-open throttle, that encode the operating context. Figure \ref{fig:hil_setup} shows a representative transformation from a raw CSV file to the corresponding prompt.

\subsection{Model selection and adaptation regime} 
Four open-source instruction-tuned families of differing scale were selected for evaluation, namely Gemma 2 2B-it, Qwen2.5 3B, Llama 3.1 8B, and Mistral 7B, on the criterion that they could follow the required structured output while remaining trainable on a single commodity GPU. Each model receives the same blind window summary and is required to return only the fault location, in the form Location\_ID: X, drawn from the six fault classes evaluated in Section 5. Restricting the response to the label in this way makes the four models directly comparable and prevents the evaluation from depending on the length or the wording of any accompanying explanation. The decision to standardise on the twenty-step window followed directly from the preliminary single-row experiments, which were unreliable because a single row does not match the temporal structure on which the models are adapted and cannot express the jumps, instability, and correlations that the descriptors encode.

\subsection{LoRA fine-tuning configuration}  
The four models are adapted with LoRA, keeping the base weights frozen and training only a small set of low-rank adapter parameters, which fits the memory budget of a single Tesla T4. Training pairs each blind signal-window prompt with its structured target, Location\_ID: X, so that the model learns to map the statistical and relational signal descriptions to one of the six predefined fault identifiers. Memory-efficient mechanisms are used throughout, i.e., low-bit quantisation of the base weights, gradient accumulation, mixed-precision computation, and gradient checkpointing where required.
The per-model settings are reported in Table 13 and are deliberately not uniform across the four families: the adapter rank ranges from one to eight, the scaling factor from two to sixteen, the maximum sequence length from 128 to 1024 tokens, and the number of epochs from five to nineteen, with a per-step batch of one, an effective batch of eight or thirty-two through gradient accumulation, and a per-model learning rate. Apart from these values, the training and evaluation logic is identical for every model, so that the comparison reflects model behaviour rather than the optimisation pipeline.

\subsection{Reproducibility and inference protocol} 

Several measures guard against label leakage. The source-file name and
the true label are excluded from the training prompt, so the model must
rely on the signal descriptors alone, and the expected output is
restricted to the location label. Partitioning is performed at the level
of independent recordings prior to windowing, so that no recording
contributes to more than one of the training, validation, and test
subsets, and a transition margin is removed around healthy--fault
boundaries so that no window spans multiple operating conditions. Each
subset is then segmented independently into sliding windows of twenty time steps in length
$W=100$ with stride $r=10$, each labelled by the class of its final time step, and Z-score normalisation statistics are computed on the
training set only and applied unchanged to the other subsets.

For reproducibility, identical preprocessing, prompt-construction,
inference, and evaluation scripts are used for every model, with adapter
weights, tokeniser settings, predictions, and training logs stored
separately from the base models. The evaluation pipeline parses the
structured Location\_ID outputs without manual correction, and
window-level predictions are aggregated by majority voting into a
file-level diagnosis, applicable both to the labelled test evaluation of
Section~5 and to blind inference on unseen recordings.

\section{Results and Discussion}
\label{sec:results}
 
The framework is evaluated on thirty-eight requirement-violating recordings from the gasoline-engine case study, distributed across six injected fault classes, after each model has been adapted with LoRA. The four models are first assessed individually, then compared globally, and the analysis closes with their inference and training cost, the interpretability of the generated diagnoses, and the resulting deployment guidance. The overall classification metrics are reported in Table~\ref{tab:overall} and the per-class breakdown in Table~\ref{tab:perclass}.
 
\begin{table*}[htbp]
\centering
\caption{Overall classification performance of the four LoRA-adapted models on the gasoline-engine test set (best value per column in bold).}
\label{tab:overall}
\begin{tabular}{lcccccc}
\toprule
Model & Accuracy (\%) & Macro-P & Macro-R & Macro-F1 & Weighted-F1 & Correct/Total \\
\midrule
Gemma 2 2B-it          & \textbf{81.58} & \textbf{0.833} & \textbf{0.817} & \textbf{0.822} & \textbf{0.818} & 31/38 \\
Qwen2.5 3B             & 76.32          & 0.783          & 0.770          & 0.764          & 0.761          & 29/38 \\
Llama 3.1 8B           & \textbf{81.58} & 0.827          & \textbf{0.817} & 0.821          & 0.817          & 31/38 \\
Mistral 7B             & 10.53          & 0.023          & 0.111          & 0.038          & 0.036          & 4/38  \\
\bottomrule
\end{tabular}
\end{table*}
 
\begin{table*}
    
\centering
\caption{Per-class precision (P), recall (R) and F1-score of the four models.}
\label{tab:perclass}
\footnotesize
\resizebox{\textwidth}{!}{%
\begin{tabular}{lccccccccccccc}
\toprule
 & & \multicolumn{3}{c}{Gemma 2 2B-it} & \multicolumn{3}{c}{Qwen2.5 3B} & \multicolumn{3}{c}{Llama 3.1 8B} & \multicolumn{3}{c}{Mistral 7B} \\
\cmidrule(lr){3-5}\cmidrule(lr){6-8}\cmidrule(lr){9-11}\cmidrule(lr){12-14}
Fault class & Sup. & P & R & F1 & P & R & F1 & P & R & F1 & P & R & F1 \\
\midrule
Acceleration pedal\_SW\_Noise   & 7 & 0.62 & 0.71 & 0.67 & 0.62 & 0.71 & 0.67 & 0.71 & 0.71 & 0.71 & 0.00 & 0.00 & 0.00 \\
Acceleration pedal\_Offset          & 6 & 0.71 & 0.83 & 0.77 & 0.67 & 0.67 & 0.67 & 0.67 & 0.67 & 0.67 & 0.00 & 0.00 & 0.00 \\
Engine\_Gain        & 6 & 1.00 & 0.83 & 0.91 & 0.83 & 0.83 & 0.83 & 1.00 & 0.83 & 0.91 & 0.00 & 0.00 & 0.00 \\
Steering wheel\_Throttle\_Noise & 6 & 0.80 & 0.67 & 0.73 & 0.71 & 0.83 & 0.77 & 0.83 & 0.83 & 0.83 & 0.14 & 0.67 & 0.23 \\
Steering wheel\_Engine\_Noise   & 7 & 0.86 & 0.86 & 0.86 & 1.00 & 0.57 & 0.73 & 0.75 & 0.86 & 0.80 & 0.00 & 0.00 & 0.00 \\
Throttle\_Noise     & 6 & 1.00 & 1.00 & 1.00 & 0.86 & 1.00 & 0.92 & 1.00 & 1.00 & 1.00 & 0.00 & 0.00 & 0.00 \\
\bottomrule
\end{tabular}%
}
\end{table*}
 
\subsection{Gemma Diagnostic Evaluation}
\label{subsec:gemma}
 
\begin{figure*}[!t]
\centering

\begin{subfigure}[t]{0.48\textwidth}
    \centering
    \includegraphics[width=0.8\linewidth]{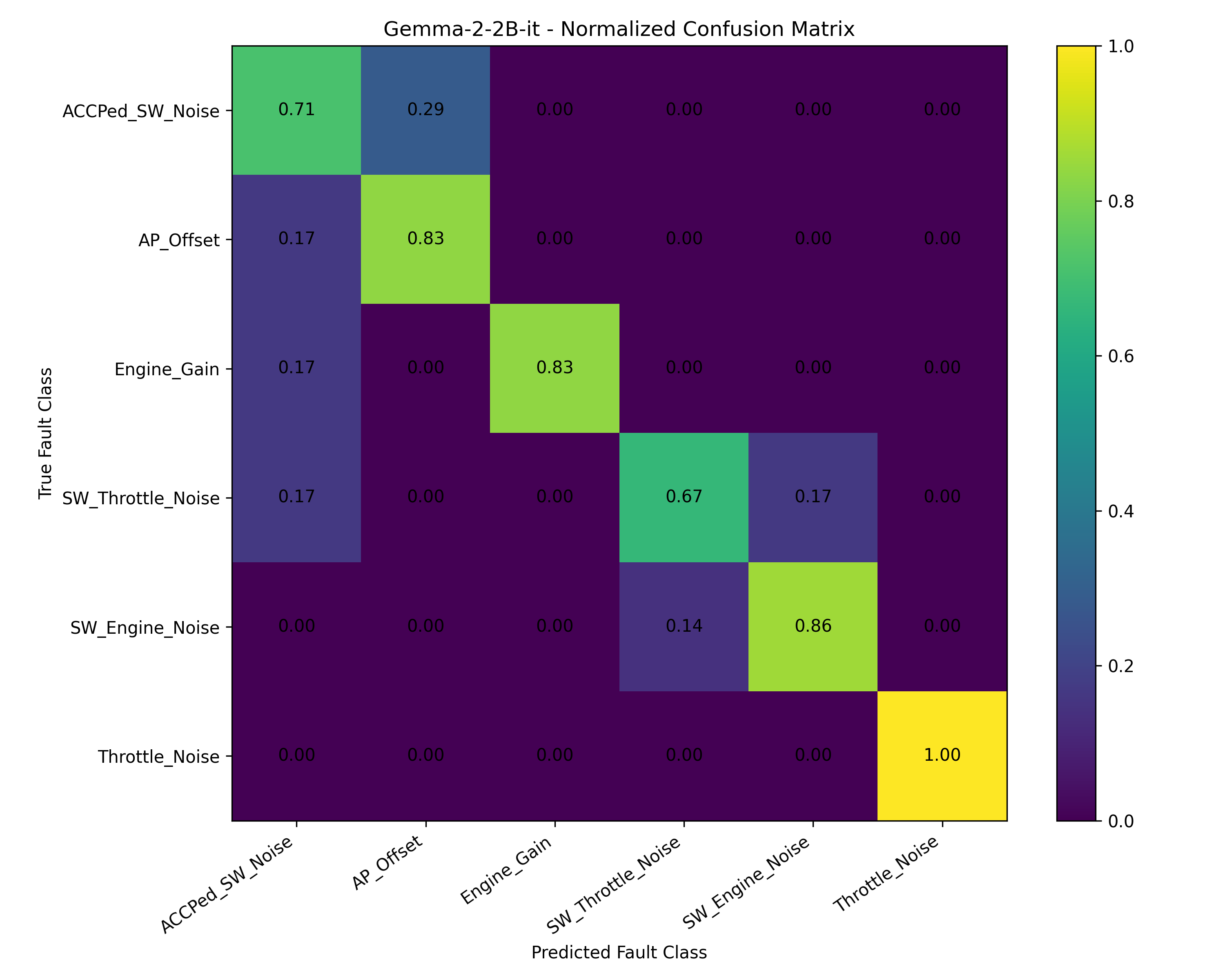}
    \caption{Gemma-2-2B-it}
    \label{fig:gemma_cm}
\end{subfigure}
\hfill
\begin{subfigure}[t]{0.48\textwidth}
    \centering
    \includegraphics[width=0.8\linewidth]{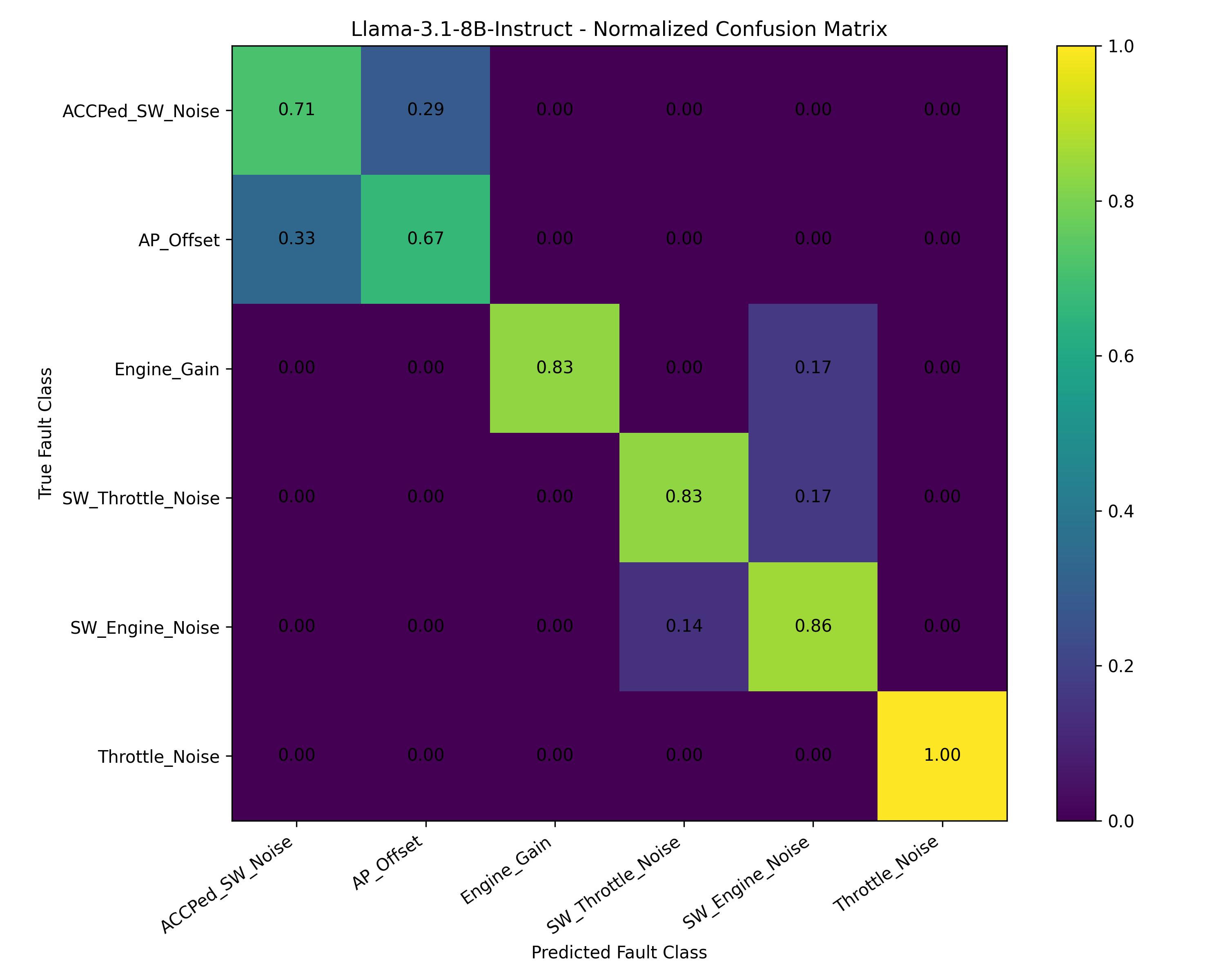}
    \caption{Llama-3.1-8B-Instruct}
    \label{fig:llama_cm}
\end{subfigure}

\vspace{0.2cm}

\begin{subfigure}[t]{0.48\textwidth}
    \centering
    \includegraphics[width=0.8\linewidth]{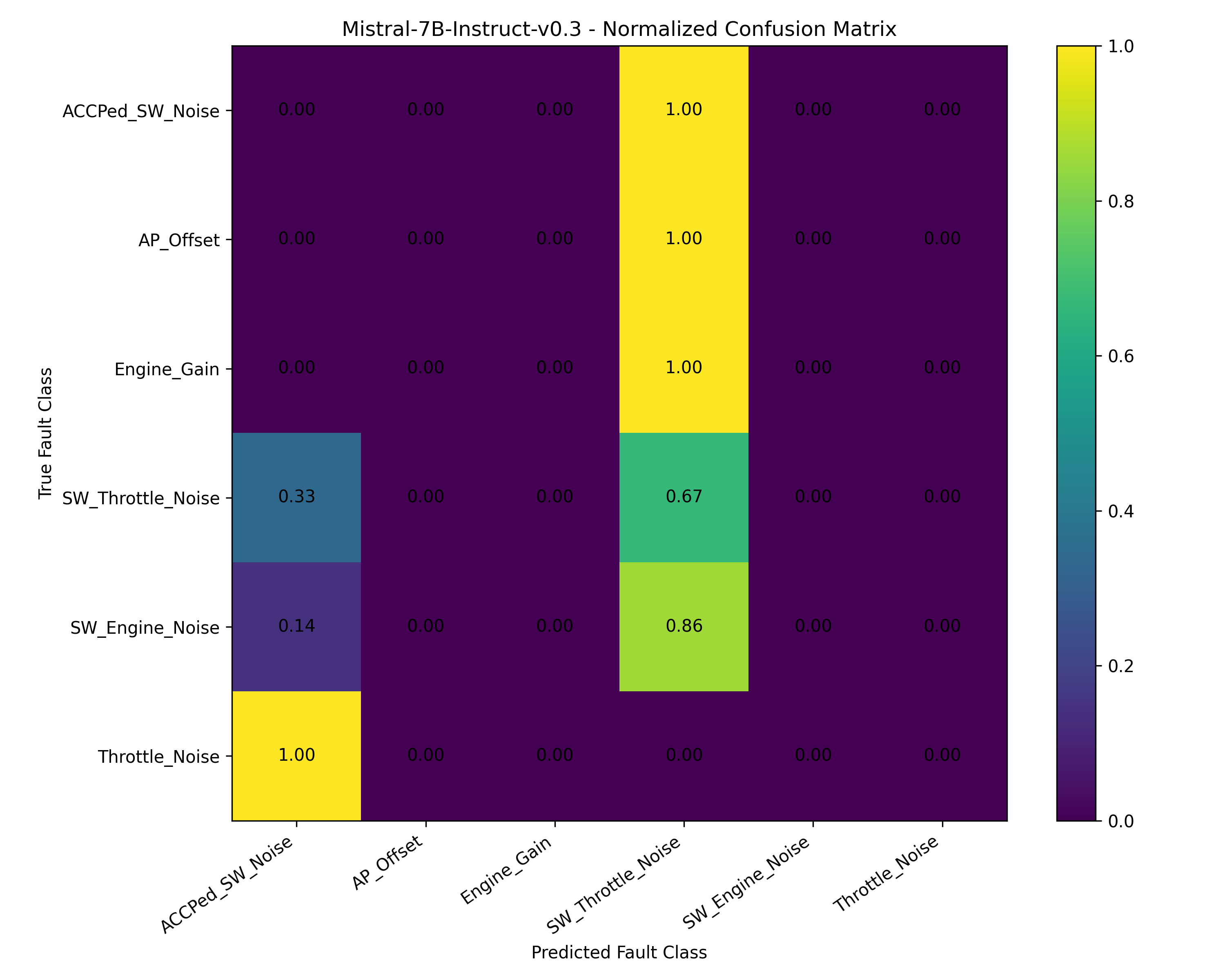}
    \caption{Mistral-7B-Instruct-v0.3}
    \label{fig:mistral_cm}
\end{subfigure}
\hfill
\begin{subfigure}[t]{0.48\textwidth}
    \centering
    \includegraphics[width=0.8\linewidth]{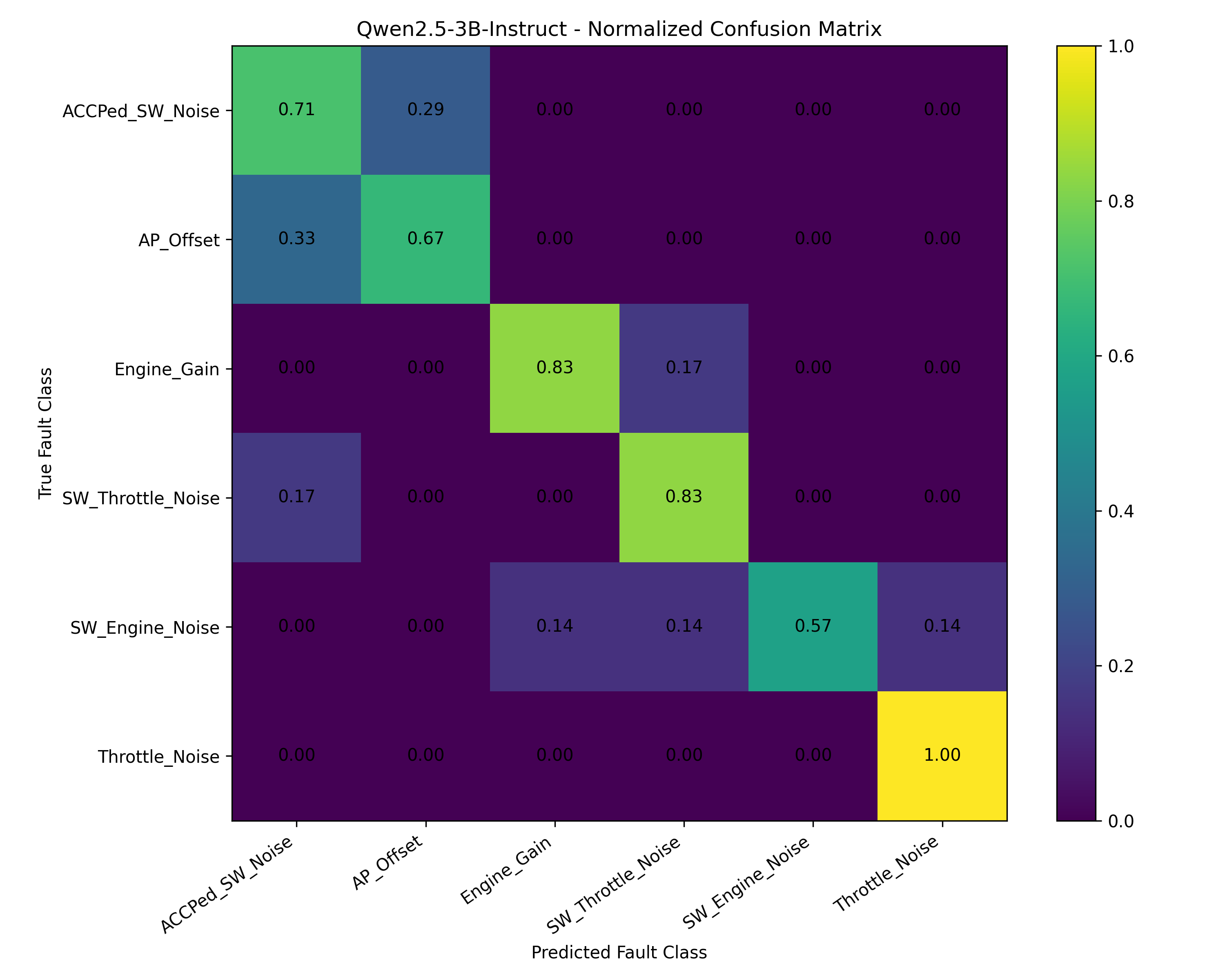}
    \caption{Qwen2.5-3B-Instruct}
    \label{fig:qwen_cm}
\end{subfigure}

\caption{Normalised confusion matrices of the evaluated LLMs for fault classification. The diagonal entries correspond to correctly classified fault classes, whereas the off-diagonal entries indicate misclassifications.}
\label{fig:llm_confusion_matrices}

\end{figure*}

Among the evaluated models, Gemma 2 2B-it correctly classified thirty-one of the thirty-eight recordings, reaching an accuracy of 81.58\% and a macro F1-score of 0.822 (Table~\ref{tab:overall}). The proximity of the macro and weighted F1 values (0.822 and 0.818) indicates that the competence of the model is distributed evenly across the six classes rather than concentrated on the better-supported ones. At the class level (Table~\ref{tab:perclass}), Throttle\_Noise is recovered without error and SW\_Engine\_Noise attains a balanced F1 of 0.86, while Engine\_Gain is isolated with perfect precision and a single missed instance. The weakest class is ACCPed\_SW\_Noise, whose precision of 0.62 stems from contamination by neighbouring predictions rather than from poor recall. As the per-model normalised confusion matrices show (Fig.~\ref{fig:gemma_cm}), the errors concentrate on the ACCPed\_SW\_Noise and AP\_Offset pair, both of which act on the accelerator-pedal path and therefore share similar statistical and relational descriptors. This behaviour can be attributed to the comparable mean shift and the similar disturbance of the pedal-to-throttle correlation produced by the two faults, which leaves little textual evidence to separate them.
 
\subsection{Qwen Diagnostic Evaluation}
\label{subsec:qwen}
 
Qwen2.5 3B operates at a slightly lower level, classifying twenty-nine of the thirty-eight recordings correctly for an accuracy of 76.32\% and a macro F1 of 0.764. The agreement between its macro and weighted F1 again indicates balanced behaviour across the classes, so the gap to the leading models is one of degree rather than of diagnostic failure. The model preserves the clean separation of Throttle\_Noise (F1 0.92) and Engine\_Gain (F1 0.83) but exhibits a distinctive weakness on SW\_Engine\_Noise, where a precision of 1.0 coincides with a recall of only 0.57. This combination is characteristic of a conservative decision boundary: the predictions that are made are correct, yet three of the seven true instances are missed and dispersed across neighbouring classes, as can be seen in Fig.~\ref{fig:qwen_cm}. The scattering of these errors, rather than their collapse onto a single class, suggests that the internal representation of this fault is insufficiently sharp rather than systematically confused with one specific alternative.
 
\subsection{Llama Diagnostic Evaluation}
\label{subsec:llama}
 
Llama 3.1 8B matches the leading performance, classifying thirty-one of the thirty-eight recordings correctly with an accuracy of 81.58\% and a macro F1 of 0.821. The two strongest models are thus separated by a factor of four in parameter count yet remain indistinguishable at this sample size, which is a first indication that scale does not govern diagnostic quality once task-specific adapters are introduced. As can be seen in Figure   Fig.~\ref{fig:llama_cm}, its per-class profile is the most balanced of the four models, holding ACCPed\_SW\_Noise at equal precision and recall and reproducing the perfect separation of Throttle\_Noise. Where Gemma is marginally stronger on SW\_Engine\_Noise, Llama is stronger on SW\_Throttle\_Noise (F1 0.83 against 0.73), so the two leaders trade minor advantages within the throttle-engine software-noise pair. The residual errors again fall on the accelerator-pedal axis, and the persistence of this confusion across both leading families confirms that the limit is set by the separability of the signals after the signal-to-text reduction rather than by model capacity.
 
\subsection{Mistral Diagnostic Evaluation}
\label{subsec:mistral}
 
Mistral 7B is the only adapted model that fails to perform the task, classifying just four of the thirty-eight recordings correctly for an accuracy of 10.53\% and a macro F1 of 0.038, close to the one-in-six level of random assignment. Five of the six classes receive an F1 of zero, and only SW\_Throttle\_Noise registers any recall. The normalised confusion matrix reveals the cause (Fig.~\ref{fig:mistral_cm}): instead of distributing its errors across plausible neighbours, the model collapses almost the entire test set onto a single label, assigning every pedal, offset and engine-gain window to SW\_Throttle\_Noise and forcing all Throttle\_Noise windows onto ACCPed\_SW\_Noise. A prediction pattern that disregards the input and defaults to one or two classes is the signature of a model that has not converged on the task, or whose output was not reliably parsed into the label schema. This outcome cannot be attributed to capacity, since Mistral is comparable in scale to the top-ranked Llama; it is most consistent with an optimisation or formatting failure specific to this model under the shared adaptation configuration. The case is retained because it marks the boundary of reliable behaviour and shows that an individual open-source model can fail silently under a regime that suits its peers.
 
\subsection{Global Performance Comparison}
\label{subsec:global}
 
\begin{figure*}
\centering
\includegraphics[width=0.8\textwidth]{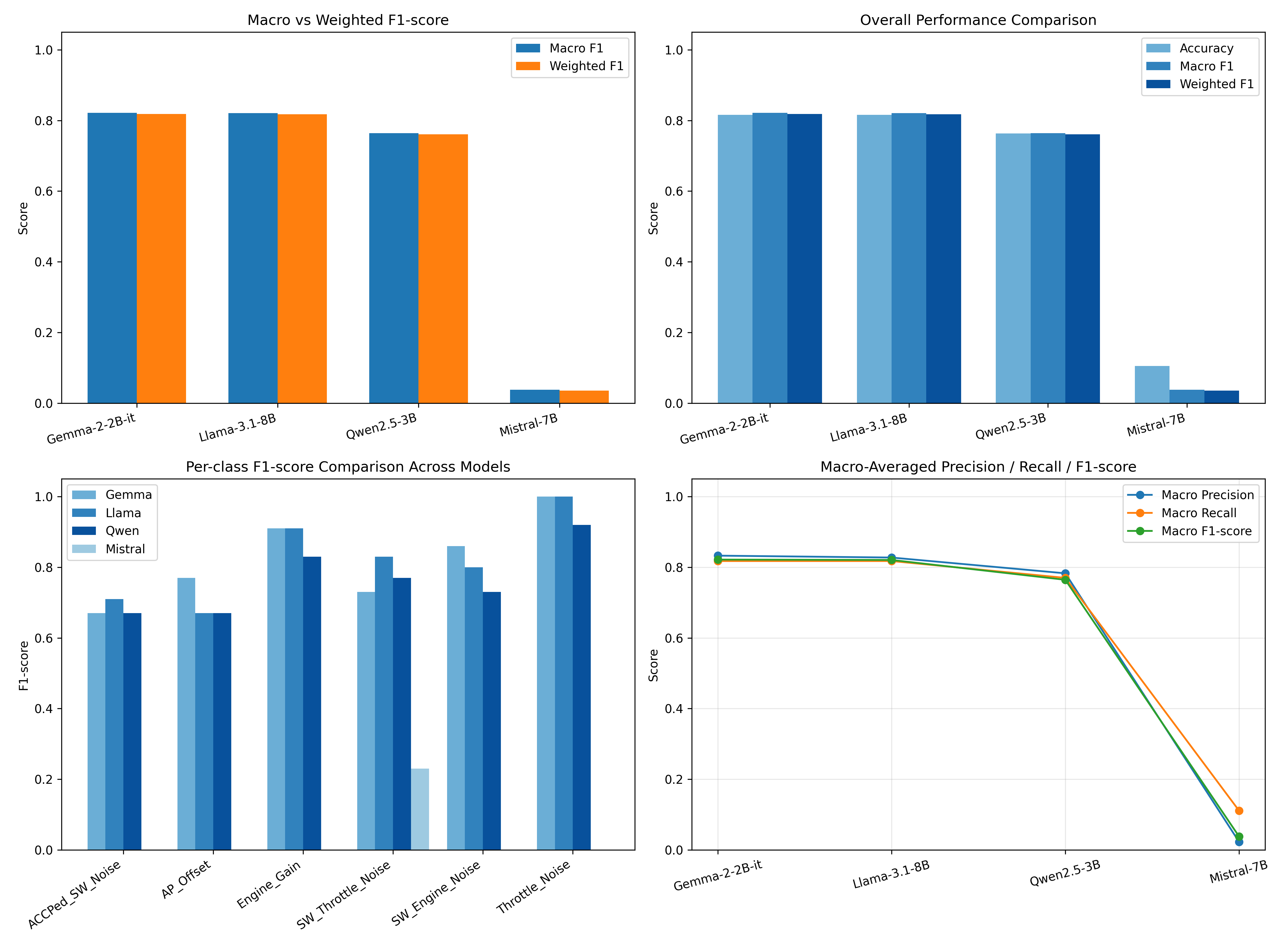}
\caption{Global performance comparison: macro versus weighted F1, overall metrics, per-class F1 across models, and macro-averaged precision, recall and F1.}
\label{fig:global_metrics}
\end{figure*}

Considered together, the four models separate into three tiers, as summarised in Fig.~\ref{fig:global_metrics}. Gemma 2 2B-it and Llama 3.1 8B lead at 81.58\% accuracy and a macro F1 of 0.82, Qwen2.5 3B follows about five points behind, and Mistral 7B stands apart through its collapse. Within each working model, accuracy, macro F1 and weighted F1 almost coincide, which confirms that the reported accuracy is not inflated by a dominant class given the near-uniform class support. The macro-averaged precision and recall curves move in step for the three successful models, indicating classifiers that are neither systematically conservative nor aggressive, whereas Mistral breaks this coupling because its over-predicted class accumulates spurious recall while precision is destroyed across the board.
 
The central observation is the decoupling of diagnostic quality from model scale. The two-billion-parameter model equals the eight-billion-parameter one, the three-billion model trails modestly, and the seven-billion model fails outright, so parameter count orders neither the successes nor the failure. Once the adapters specialise each base model to the descriptor-to-fault mapping, performance depends on how well a given family converges on this narrow task rather than on its raw capacity. The per-class comparison in Fig.~\ref{fig:global_metrics} and Table~\ref{tab:perclass} confirms a shared difficulty profile across the working models, in which Throttle\_Noise and Engine\_Gain are recovered reliably, the accelerator-pedal pair remains the hardest distinction, and the throttle-engine software-noise pair forms the second source of error. The conservation of this profile across three independent families locates the residual difficulty in the data rather than in any single model.
 

\subsection{Computational Cost Analysis of Inference Testing and Training}
\label{subsec:cost}
 
The computational cost of inference and training is reported in Table~\ref{tab:inference_cost} and Table~\ref{tab:training_cost}. Inference latency scales with model size: Gemma and Qwen process the test set in close to two minutes, at about 3.2 and 3.4 seconds per sample, whereas Llama and Mistral require between 5.6 and 6.6 minutes, at 8.9 and 10.4 seconds per sample, with peak reserved memory rising correspondingly from roughly 1.4 GB for Gemma, the lightest model, to 5.2 GB for Mistral. Training time does not follow parameter count monotonically, because total time combines the per-epoch cost with the number of epochs needed for convergence; Llama is the cheapest per epoch yet the most expensive overall, while Mistral carries the highest per-epoch cost and still fails to converge, the most demanding training producing the weakest classifier. Peak training memory stays between 5.3 and 9.9 GB allocated, which keeps the entire adapt-and-evaluate cycle within the sixteen-gigabyte budget of a single Tesla T4.
 
\begin{table*}
\centering
\caption{Computational cost of inference testing for the evaluated models.}
\label{tab:inference_cost}
\footnotesize
\resizebox{\textwidth}{!}{%
\begin{tabular}{lcccccccc}
\toprule
Model & Test & Wall & Wall & Seconds & Samples & Peak alloc. & Peak res. & GPU \\
      & samples & time (s) & time (min) & /sample & /second & (MB) & (MB) & \\
\midrule
Gemma 2 2B-it          & 38 & 122.020 & 2.034 & 3.2111 & 0.3114 & 12.19   & 14.00   & Tesla T4 \\
Qwen2.5 3B             & 38 & 130.741 & 2.179 & 3.4406 & 0.2907 & 1603.46 & 1630.00 & Tesla T4 \\
Llama 3.1 8B           & 38 & 338.343 & 5.639 & 8.9038 & 0.1123 & 3128.36 & 3176.00 & Tesla T4 \\
Mistral 7B             & 38 & 394.197 & 6.570 & 10.3736 & 0.0964 & 5128.25 & 5208.00 & Tesla T4 \\
\bottomrule
\end{tabular}%
}
\end{table*}
 
\begin{table*}
\centering
\footnotesize
\caption{Training computational cost of the fine-tuned models.}
\label{tab:training_cost}
\begin{tabular}{lcccc}
\toprule
Model & Time (min) & Seconds/epoch & Peak alloc. GPU (GB) & Peak reserved GPU (GB) \\
\midrule
Gemma 2 2B-it & 32.16  & 385.88 & 9.88 & 10.97 \\
Qwen2.5 3B    & 53.91  & 539.06 & 5.27 & 6.08  \\
Llama 3.1 8B  & 110.63 & 348.77 & 9.69 & 10.79 \\
Mistral 7B    & 76.58  & 918.92 & 9.30 & 9.61  \\
\bottomrule
\end{tabular}
\end{table*}

\begin{table*}[H]
\centering
\caption{Practical interpretation of inference efficiency and diagnostic suitability.}
\label{tab:practical}
\footnotesize
\renewcommand{\arraystretch}{1.2}
\begin{tabular}{p{2.2cm} p{2.6cm} p{2.6cm} p{4.2cm} p{2.0cm}}
\toprule
Model & Inference efficiency & Diagnostic performance & Practical interpretation & Overall suitability \\
\midrule
Gemma 2 2B-it &
Fastest model with the lowest inference time per sample &
Strong performance with 81.58\% accuracy and the highest weighted F1-score &
Best balance between speed, memory efficiency, and diagnostic accuracy. Suitable for resource-constrained diagnostic workflows. &
Highly suitable \\
\addlinespace
Qwen2.5 3B &
Fast and only slightly slower than Gemma &
Good performance with 76.32\% accuracy &
Good compromise between runtime and classification quality. Useful when stable instruction-following behaviour is required, but slightly weaker than Gemma and Llama. &
Suitable \\
\addlinespace
Llama 3.1 8B &
Slower and more computationally demanding &
Strong performance with 81.58\% accuracy, similar to Gemma &
Provides strong diagnostic accuracy but requires substantially more inference time and GPU memory. Suitable when higher computational cost is acceptable. &
Suitable but costly \\
\addlinespace
Mistral 7B &
Slowest model with the highest computational cost &
Weak performance with 10.53\% accuracy &
Least practical in the current setup due to high runtime, high memory usage, and low classification reliability. Requires further optimisation. &
Not suitable in current setup \\
\bottomrule
\end{tabular}
\end{table*}

\section{Conclusion}
\label{sec:conclusion}

Validating safety-critical automotive software produces more recordings than manual review can handle, yet ISO 26262 requires each deviation to be detected, characterised, located and justified. Threshold checks stop at detection, and deep-learning classifiers need abundant labelled data while hiding their reasoning. This article asks whether open-source instruction-tuned language models, given a compact textual account of sensor behaviour, can deliver data-efficient diagnoses open to engineering scrutiny.
The framework separates validation from diagnosis. Requirement checking on a dSPACE real-time platform retains only the recordings that violate a safety requirement; each retained window is reduced to statistical, relational, and contextual descriptors, written into a fixed prompt, and mapped by a quantised low-rank-adapted model to a fault location with an explanation citing the descriptor evidence.
On a gasoline-engine case study with six fault classes, the two-billion-parameter Gemma matched the eight-billion-parameter Llama at 81.6\% accuracy and a macro F1 near 0.82; Qwen trailed by a few points, and Mistral collapsed despite comparable scale, decoupling diagnostic quality from parameter count. Errors were confined to physically adjacent faults, so explanations stayed plausible even when a label was wrong, and the full adapt-and-evaluate cycle ran offline within sixteen gigabytes on a single commodity accelerator, making the smallest model the most practical choice.

The single powertrain and modest recording set make the scores indicative rather than definitive; one model's collapse argues for per-model tuning and stability checks, and the adjacent-fault confusion reflects information discarded by the signal-to-text reduction. Future work will enrich the descriptors, broaden the fault catalogue and operating conditions, evaluate severity assessment, and study transfer to other propulsion architectures.

\bibliographystyle{IEEEtran}
\bibliography{Reference}

\end{document}